\documentclass[12pt]{article}
\usepackage{graphicx}
\newlength{\vshift}
\input epsf.tex
\newlength{\hshift}

\usepackage{amssymb}
\usepackage{amsmath}
\usepackage{amsthm}
\usepackage{amsopn}
\begin{document}

 \vspace*{3cm}

\begin{center}

{\bf{\Large    Lorentz Violation in the Higgs Sector and
Noncommutative Standard Model}}

\vskip 4em

{ {\bf S. Aghababaei$^*$} \footnote{e-mail:
s.aghababaei@ph.iut.ac.ir}\: , \: {\bf M.
Haghighat$^*$}\footnote{e-mail:mansour@cc.iut.ac.ir}\\ and
\\
 {\bf A. Kheirandish$^{*\dagger}$}\footnote{e-mail:ali.kheirandish@icecube.wisc.edu}}
\vskip 1em

$^*$Department of Physics, Isfahan University of Technology,
Isfahan, 84156-83111, Iran\\
$^\dagger$Wisconsin IceCube Particle Astrophysics Center (WIPAC),
University of Wisconsin, Madison, Wisconsin 53703, USA.

 \end{center}

 \vspace*{1.9cm}

\begin{abstract}
The noncommutative standard model apparently violates the Lorentz
symmetry.  We compare the Lorentz violating terms in the Higgs
sector of the noncommutative standard model with their counterparts
in the standard model extension.  We show that the Lorentz violating
parameters in the Higgs sector can be expressed directly in terms of
the noncommutative parameter without any background field.  The
absence of the background field enhances the obtained bounds on the
noncommutative parameter from the standard model extension.
\end{abstract}

\newpage
\section{Introduction}
Although the  Lorentz symmetry is a well tested symmetry of nature,
the possibility that new physics involves a Lorentz symmetry
violation  has been considered in many works.  The main motivation
to consider such a violation is in the fact that in the Planck
scale, where the quantum gravity should be considered, the Lorentz
symmetry violation arises naturally.  V. A. Kostelecky and S. Samuel
\cite{ssb} showed that the Lorentz symmetry can be broken
spontaneously in the context of string theory.  Consequently, D.
Colladay and V. Alan Kostelecky \cite{SME}, irrespective of the
underlying fundamental theory, introduced a general extension of the
minimal standard model that violates both Lorentz invariance and
CPT.  The phenomenological aspects of the so called Standard Model
Extension (SME) have been extensively considered by many authors in
terrestrial \cite{terrestrial} and astrophysical systems
\cite{astro}, and the bounds on the Lorentz Violating parameters
(LV) are collected in \cite{table}.  Meanwhile, in noncommutative
(NC) space-time, where in its canonical version, the coordinates are
operators and satisfy the relation:
\begin{equation}
    [\hat{x}_\mu\, , \hat{x}_\nu]=\theta_{\mu\nu}=\frac{\varepsilon_{\mu\nu}}{\Lambda^2_{NC}},\label{NC}
\end{equation}
the real, constant, and antisymmetric parameter $\theta_{\mu\nu}$
breaks the Lorentz symmetry intrinsically. Therefore, the standard
model in noncommutative space-time may be considered as a subset of
the SME.  There are two approaches to construct the gauge theories
and consequently, the standard model in noncommutative space. In the
first one, the gauge group is restricted to $U(n)$, and the symmetry
group of the standard model is achieved by the reduction of $U(3)
\times U(2) \times U(1)$ to $SU(3) \times SU(2) \times U(1)$ by an
appropriate symmetry breaking \cite{NCSM1}. In the second one, the
noncommutative gauge theory can be constructed for a $SU(n)$ gauge
group via a Seiberg-Witten map \cite{SW} where the fields
themselves, in contrast with the first approach,  depend on the
parameter of non-commutativity \cite{NCSM2}. However, the NC-field
theories and their phenomenological aspects based on both versions
have been examined for many years \cite{NCpheno}-\cite{ncth}. The
relation between the NC-field theory with the $U(1)$ gauge group
based on both approaches is compared with the QED part of the SME in
\cite{NC-SME}. In both versions, the Lorentz violating parameters
depend on the NC-parameter through a magnetic field as a background;
and in the absence of the background, all the LV-parameters are
zero. In this article, we would like to explore the relation between
the Higgs part of full SME and the noncommutative standard model
(NCSM) based on the second approach to find those explicit relations
between NC and LV-parameters without any axillary fields.

\noindent In section 2 we introduce the Higgs sector of the NCSM
based on the $SU(3)\times SU(2) \times U(1)$ gauge group and its
counterpart in the Lorentz violating extension of the standard
model. In Section 3 we derive the LV-parameters in terms of the
NC-parameter in the absence of the background field. The bounds on
the noncommutative scale and some concluding remarks are given in
section 4.

\section{the Higgs part of SME and NCSM}
The SME is an extension of the standard model in which all possible
Lorentz violating terms that could arise from the spontaneous
symmetry breaking at a fundamental level are included and should
preserve the gauge symmetry $SU(3)\times SU(2) \times U(1)$ with a
power counting renormalizabilty.  Therefore, the Higgs part of the
standard model
\begin{eqnarray}
\mathcal{L}_{Higgs}=(D_{\mu}\phi)^{\dag} D^{\mu}\phi-\mu^2
\phi^{\dag} \phi -\lambda (\phi^{\dag} \phi)^2,\label{SMhiggs}
\end{eqnarray}
where the covariant derivative with the field strength  $B_{\mu\nu}$
for the hypercharge and $W_{\mu\nu}$ for $SU(2)$ is defined as
\begin{eqnarray}
D_{\mu}=\partial_\mu+ig'\frac{Y}{2}B_{\mu\nu}+igT_aW^a_{\mu\nu},
\end{eqnarray}
can be extended to \cite{SME}
\begin{eqnarray}
\mathcal{L}_{Higgs}+\mathcal{L}^{CPT-even}_{Higgs}+\mathcal{L}^{CPT-odd}_{Higgs},
\end{eqnarray}
where
\begin{eqnarray}\label{cpteven}
\mathcal{L}^{CPT-even}_{Higgs}&=&(\frac{1}{2}(k_{\phi\phi})^{\mu\nu}(
D_{\mu} \phi^{\dag}) D_{\nu}\phi+ h.c.)\\\nonumber
&-&\frac{1}{2}(k_{\phi B})^{\mu\nu} \phi^{\dag} \phi
B_{\mu\nu}-\frac{1}{2}(k_{\phi W})^{\mu\nu} \phi^{\dag} W_{\mu\nu}
\phi,
\end{eqnarray}
indicates the CPT preserving part of the Lagrangian, and
\begin{eqnarray}
\mathcal{L}^{CPT-odd}_{Higgs}=i (k_{\phi})^{\mu}\phi^{\dag}
D_{\mu}\phi +h.c.\,,\label{cptodd}
\end{eqnarray}
which is odd under the CPT-symmetry.

In NC-space-time, the coordinates are operators, and in the
canonical version they satisfy (\ref{NC}). To construct the
noncommutative field theory, according to the Weyl-Moyal
correspondence, an ordinary function can be used instead of the
corresponding noncommutative one by replacing the ordinary product
with the star product, as in
\begin{eqnarray}
(f \star g)(x)= \left.\exp\!\left(\frac{i}{2}
     \theta^{\mu \nu}\frac{\partial}{\partial x^\mu}
     \frac{\partial}{\partial y^\nu}\right) f(x) g(y)\right|_{y \to x},
\end{eqnarray}

where up to the first order of  $\theta^{\mu \nu}$,

\begin{eqnarray}
f \star g = f \cdot g + \frac{i}{2}\theta^{\mu\nu}\partial_\mu f
\cdot \partial_\nu g + \mathcal{O}(\theta^2) .
\end{eqnarray}
Using this correspondence is not enough to construct a gauge theory.
 Only a $U(n)$ gauge theory with some restriction on the allowed representations
 can be simply extended to the noncommutative gauge theory.  However, a minimal
 way to get rid of these restrictions is the construction of a gauge theory with a $SU(n)$
 gauge group via the Seiberg-Witten map which provides the noncommutative
 fields as local functions of the ordinary fields \cite{NCSM2}.
 Therefore, to construct the NCSM, we should replace the ordinary
 products and fields in the ordinary standard model with the star
 product and NC-fields, respectively.  Consequently, the action for the
  Higgs part of the NCSM can be easily constructed from
 (\ref{SMhiggs}) as follows

\begin{eqnarray}\label{action}
S_{Higgs-NCSM}&=&\int d^4x \bigg( \rho_0(\widehat D_\mu \widehat
\Phi)^\dagger \star \rho_0(\widehat D^\mu \widehat \Phi) - \mu^2
\rho_0(\widehat {\Phi})^\dagger \star  \rho_0(\widehat
\Phi)\\\nonumber &-& \lambda \rho_0(\widehat \Phi)^\dagger \star
\rho_0(\widehat \Phi) \star \rho_0(\widehat \Phi)^\dagger \star
\rho_0(\widehat \Phi) \bigg),
\end{eqnarray}
where the hat shows the noncommutative field and $\rho_0$ realizes
an appropriate representation for the hybrid Seiberg-Witten map. The
covariant derivative is defined as
\begin{eqnarray}
D_{\mu}=\partial_\mu+iV_\mu,
\end{eqnarray}
where the gauge potential $V_{\mu}$ is defined as

\begin{equation}
 {V_\mu}=g' B_\mu(x)\frac{Y}{2}+g \sum_{a=1}^{3} W_{\mu a}(x) T^a_L,
\end{equation}
in which $Y$ and $T^{a}_{L}$ are the generators of $U(1)_Y$ and
$SU(2)_L$, respectively.

Now the Higgs action (\ref{action}) can be expanded to all orders of
$\theta^{\mu \nu}$.  For this purpose one needs the $\theta^{\mu
\nu}$-dependent of the gauge and Higgs fields to all orders. Up to
the first order of the NC-parameter, one has
\begin{equation}
    \rho_0(\hat \Phi)=\phi+\rho_0(\phi^1)+\mathcal{O}(\theta^2),
    \end{equation}
 for the Higgs field with
\begin{eqnarray}
      \rho_0(\phi^1)=-\frac{1}{2}\theta^{\alpha \beta}
      (g'B_\alpha+g W_\alpha) \partial_\beta \phi
      +i\frac{1}{4} \theta^{\alpha \beta}
      (g'B_\alpha+g W_\alpha) (g'B_\beta+g W_\beta) \phi,
\end{eqnarray}
where $W_\alpha=W_\alpha^a T_a$.
 Meanwhile, the expansion for the mathematical field $V$ up to
the leading order is given by
\begin{eqnarray}
\widehat V_\mu=V_\mu+ i\Gamma_\mu + {\cal O}(\theta^2),
\end{eqnarray}
\noindent with
\begin{eqnarray}
\Gamma_\mu & = &   i\frac{1}{4}\theta^{\alpha \beta}
     \{ g' B_\alpha + g W_\alpha \,,\,
     g' \partial_\beta B_\mu + g \partial_\beta W_\mu
     + g' B_{\beta \mu} +g W_{\beta \mu} \}.
\end{eqnarray}

\noindent where $B_{\mu\nu}$ and $W_{\mu\nu}$ are the ordinary field
strengths for the hypercharge and the $SU(2)$ gauge fields. However,
one should note that the NC-field strength has the following
expansion
\begin{eqnarray}
  \widehat F_{\mu \nu}&=&F_{\mu \nu}+ F^1_{\mu \nu} +{\cal O}(\theta^2),
\end{eqnarray}
\noindent with
\begin{eqnarray}
  F_{\mu \nu}&=&g'B_{\mu \nu}+g W_{\mu \nu},
\end{eqnarray}
 and
\begin{eqnarray}
  F^1_{\mu \nu}&=& \frac{1}{2} \theta^{\alpha \beta} \{ F_{\mu \alpha},
  F_{\nu \beta} \} -\frac{1}{4} \theta^{\alpha \beta}
  \{ V_\alpha,(\partial_\beta+D_\beta) F_{\mu \nu} \}.
\end{eqnarray}

Therefore, the Higgs action up to the first order of the
NC-parameter results in

\begin{eqnarray}\label{SHiggs}
   S_{Higgs}&=& \int d^4x\Bigg( (D^{SM}_\mu\phi)^\dagger D^{SM \mu}\phi
   -\mu^2 \phi^\dagger \phi
-\lambda (\phi^\dagger \phi) (\phi^\dagger \phi) \Bigg)
   \\ \nonumber &&
   +
   \int d^4x \Bigg ( (D^{SM}_\mu\phi)^\dagger
   \left( D^{SM \mu}\rho_0(\phi^1) + \frac{1}{2}
   \theta^{\alpha \beta} \partial_\alpha V^{\mu} \partial_\beta \phi
 + \Gamma^\mu \phi \right)
\\ \nonumber && +
\left(D^{SM}_\mu \rho_0 (\phi^1) + \frac{1}{2}
   \theta^{\alpha \beta} \partial_\alpha V_\mu \partial_\beta \phi
 + {\Gamma_\mu} \phi \right)^\dagger D^{SM \mu}\phi
\\ \nonumber &&
+\frac{1}{4} \mu^2 \theta^{\mu \nu} \phi^\dagger (g' B_{\mu \nu} + g
W_{\mu \nu}) \phi -  \lambda i \theta^{\alpha \beta} \phi^\dagger
\phi (D^{SM}_\alpha \phi)^\dagger (D^{SM}_\beta \phi) \Bigg) + {\cal
O}(\theta^2).
  \end{eqnarray}

\section{Lorentz Violating Coefficients}
Noncommutative coordinates (\ref{NC}) apparently violate the Lorentz
symmetry.  Therefore, the Higgs action given in (\ref{SHiggs})
violates the symmetry too.  In previous works \cite{NC-SME}, the
NC-parameter was related to its correspondence in the SME ( in fact,
to the QED part of the SME ) through a magnetic-field as a
background.   Here we are looking for the direct relation between
the parameters of both theories.  To this end, we compare
(\ref{SHiggs}) with (\ref{cpteven}) and (\ref{cptodd}) regarding the
absence of background.  One can easily find

\begin{eqnarray}
(k_{\phi\phi})^{\mu\nu}&=&-2i\lambda \phi^{\dag}
\phi\theta^{\mu\nu}+(K_{\phi\phi})^{\mu\nu}(B,W),
\end{eqnarray}
where $K_{\phi\phi}$ stands for the gauge field dependent part of
$k_{\phi\phi}$, which is zero in the absence of background.  It
should be noted that in general, $k_{\phi\phi}$ has symmetric and
antisymmetric parts, as

\begin{eqnarray}
(k_{\phi\phi})^{\mu\nu}=(k^S_{\phi\phi}+ik^A_{\phi\phi})^{\mu\nu}.
\end{eqnarray}
Therefore, in the Higgs part of the NCSM, only the antisymmetric
part of $k_{\phi\phi}$ is nonzero, where after the symmetry breaking
one has
\begin{eqnarray}
(k^A_{\phi\phi})^{\mu\nu}&=&-2\lambda
(\frac{v}{\Lambda_{NC}})^2\varepsilon^{\mu\nu}=-(\frac{M_H}{\Lambda_{NC}})^2\varepsilon^{\mu\nu}.\label{k1}
\end{eqnarray}
As is expected for the CPT conserved NC-field theory,
$(k_{\phi})_{\mu}=0$  in (\ref{SHiggs}).  Comparing (\ref{SHiggs})
with (\ref{cpteven}) for the LV-parameters $(k_{\phi B})_{\mu\nu}$
and $(k_{\phi W})_{\mu\nu}$ leads to

\begin{eqnarray}
(k_{\phi B})^{\mu\nu}=-\frac{1}{2}\lambda  g'
(\frac{v}{\Lambda_{NC}})^2\varepsilon^{\mu\nu}=-\frac{e}{4\cos\theta_W}
(\frac{M_H}{\Lambda_{NC}})^2\varepsilon^{\mu\nu},\label{k2}
\end{eqnarray}
and
\begin{eqnarray}
(k_{\phi W})^{\mu\nu}=-\frac{1}{2}\lambda g
(\frac{v}{\Lambda_{NC}})^2\varepsilon^{\mu\nu}=-\frac{e}{4\sin\theta_W}
(\frac{M_H}{\Lambda_{NC}})^2\varepsilon^{\mu\nu}.\label{k3}
\end{eqnarray}
The value $\frac{|k_{\phi B}|}{|k_{\phi W}|}=\tan\theta_W$ is in
agreement with the experimental values given in Table{\ref{Higgs}}.
In fact, in \cite{exp-higgs} the experimental bounds on the
LV-parameters in the Higgs sector are indirectly obtained by
evaluating the photon vacuum polarization at one-loop and comparing
the obtained results with the experimental bounds on the $k_F$-term.
As (\ref{cpteven}) shows, the photon-Higgs coupling in the
hypercharge term is $\sim\cos\theta_W$, while in the $W^3$-term it
is $\sim\sin\theta_W$, which leads to $(\frac{|k_{\phi B}|}{|k_{\phi
W}|})_{Exp.}=\tan\theta_W$.  One should note that in
(\ref{k1})-(\ref{k3}) there are subdominant terms depending on
$\frac{eB}{M_H^2}$ that are very small, even for a magnetic field as
large as $10^{13}$ teslas. Therefore, even in the prepense of the
magnetic field, only the B-independent terms are enough to find a
bound as large as $10^6 TeV$ for the NC-parameter. Such a bound on
the $\Lambda_{NC}$ is too large, compared with the bounds of $ GeV $
in low energy experiments \cite{NCpheno} and a few $TeV$ in high
energy scattering \cite{1tev}.


\section{Conclusion}
We examined the Higgs sector of the NCSM based on the $SU(3)\times
SU(2) \times U(1)$ gauge symmetry.  As a subset of the standard
model extension, we compared the Higgs sectors in the both theories.
We found the LV-parameters $k_{\phi B}$, $k_{\phi W}$, and
$k_{\phi\phi}$ for the Higgs sector as a function of the
NC-parameter ( see (\ref{k1})-(\ref{k3})).  In the NCSM, only the
antisymmetric part of $k_{\phi\phi}$ survives in the absence of an
electromagnetic background field.  For all the LV-parameters in the
NC-space, there are also corrections of an order of
$\frac{eB}{M_H^2}$ smaller than the background independent part,
which is too small to be considered here. In the previous works to
relate the parameters of the both theories, one needs a constant
magnetic background and, in the absence of the background, the LV
and the NC-parameters decoupled from each other \cite{NC-SME}. As
the obtained results show, here is the first place in which the
noncommutativity  is directly expressed in terms of the
LV-parameters. The experimental bounds on the antisymmetric parts of
$k_{\phi\phi}$, $k_{\phi B}$, and $k_{\phi W}$ leads to a bound on
the NC-parameter as large as $\Lambda_{NC}\sim 10^6 \,TeV$, see
Table{\ref{Higgs}}. In fact the main result is twofold : 1-Direct
relation between the LV and NC parameters without any axillary
field. 2- A very large bound on the NC-parameter compared to the
current bound of the order of a few $TeV$.

\begin{table}[h]
    \begin{center}
\begin{tabular}{|c|c|c|c|c|}
  \hline
  Coefficient& NC-expression & Exp.\cite{table}\cite{exp-higgs} & System & $\Lambda_{NC}(TeV)$  \\
  \hline
  $k^A_{\phi\phi}$ & $-(\frac{M_H}{\Lambda_{NC}})^2$&$3\times 10^{-16}$ & Cosmological & $7\times 10^6$    \\
  $k_{\phi B}$ & $-\frac{e}{4\cos\theta_W}
(\frac{M_H}{\Lambda_{NC}})^2$ &$0.9\times 10^{-16}$ & Cosmological & $4\times 10^6$    \\
  $k_{\phi W}$ &$-\frac{e}{4\sin\theta_W}
(\frac{M_H}{\Lambda_{NC}})^2$ &$1.7\times 10^{-16}$ & Cosmological &  $4\times 10^6$   \\

  \hline
  \end{tabular}
\end{center}
    \caption{ Higgs Sector LV-Coefficients}\label{Higgs}
    \end{table}


\begin{thebibliography}{99}
\bibitem{ssb}V. A. Kostelecky, and S. Samuel, Phys. Rev. Lett. {\bf 63}, 224 (1989); Phys. Rev. D {\bf 39}, 683
(1989).

\bibitem{SME}
D. Colladay and V. A. Kostelecky,  Phys. Rev. D {\bf 58}, 116002
(1998); Phys. Rev. D {\bf 55}, 6760 (1997).
\bibitem{terrestrial}
O.M. Del Cima, D.H.T. Franco, A.H. Gomes, J.M. Fonseca and O.
Piguet, Phys. Rev. D {\bf 85}, 065023 (2012); D. Marfatia, H. Pas,
S. Pakvasa and T. J. Weiler, Phys. Lett. B {\bf 707}, 553 (2012), D.
Colladay and P. McDonald, Phys. Rev. D {\bf 85}, 044042 (2012); M.
Cambiaso, R. Lehnert and R. Potting, Phys. Rev. D {\bf 85}, 085023
(2012); MINOS Collaboration (P. Adamson, et al.), Phys. Rev. D {\bf
85}, 031101 (2012); R. Potting, Phys. Rev. D {\bf 85}, 045033
(2012); V. A. Kostelecky and M. Mewes, Phys. Rev. D {\bf 85}, 096005
(2012); F. Bezrukov and H. M. Lee, Phys. Rev. D {\bf 85}, 031901
(2012); V. Baccetti, K. Tate and M. Visser, JHEP {\bf 1203}, 087
(2012); B. Altschul, Phys. Rev. D {\bf 84}, 091902 (2011); Chun Liu,
Jin-tao Tian and Zhen-hua Zhao, Phys. Lett. B {\bf 702}, 154 (2011);
B. Altschul, Phys. Rev. D {\bf 84}, 076006 (2011); J. S. Diaz and V.
A. Kostelecky,
 Phys. Rev. D {\bf 85}, 016013 (2012); Jorge S. Diaz and Alan Kostelecky, Phys. Lett. B {\bf 700}, 25 (2011).

\bibitem{astro}
I. Motie and She-Sheng Xue, Int. J. Mod. Phys. A {\bf 27}, 1250104
(2012); D. Hernandez and M. Sher, Phys. Lett. B {\bf 698}, 403
(2011); L. Shao and Bo-Qiang Ma, Phys. Rev. D {\bf 83}, 127702
(2011); M. A. Hohensee, S. Chu, A. Peters and H. Muller, Phys. Rev.
Lett. {\bf 106}, 151102 (2011); F. W. Stecker, Astropart. Phys. {\bf
35}, 95 (2011); L.C. Garcia de Andrade, Phys. Lett. B {\bf 696}, 1
(2011); G. Cacciapaglia, A. Deandrea and L. Panizzi, JHEP {\bf 11},
137 (2011); A. Saveliev, L. Maccione and G. Sigl, JCAP {\bf 1103},
046 (2011); IceCube Collaboration (R. Abbasi, et al.), Phys. Rev. D
{\bf 82}, 112003 (2010);
 M.~Zarei, E.~Bavarsad, M.~Haghighat, I.~Motie,
R.~Mohammadi, Z.~Rezaei, Phys. Rev. D 81, 084035 (2010); L.
Campanelli and P. Cea, Phys. Lett. B {\bf 675}, 155 (2009).

\bibitem{table}
A. Kostelecky and N. Russell, Rev. Mod. Phys. {\bf 83},11(2011).

\bibitem{NCSM1}
M. Chaichian, P. Presnajder, M.M. Sheikh-Jabbari, and A. Tureanu,
Eur. Phys. J. C {\bf 29}, 413 (2003).

\bibitem{SW}
N. Seiberg and E. Witten, J. High Energy Phys. {\bf 09}, 032 (1999).

\bibitem{NCSM2}
X. Calmet, B. Jurc¡o, P. Schupp, J. Wess, and M. Wohlgenannt, Eur.
Phys. J. C {\bf 23}, 363 (2002); B. Melic´, K. Passek-Kumeric¡ki, J.
Trampetic´, P. Schupp, and M. Wohlgenannt, Eur. Phys. J. C {\bf 42},
483 (2005).

\bibitem{NCpheno}
M. Chaichian, P. Presnajder, M. M. Sheikh-Jabbari, and A. Tureanu,
Phys. Lett. B {\bf  683}, 55 (2010); M. Zarei, E. Bavarsad, M.
Haghighat, I. Motie, R. Mohammadi and Z. Rezaei, Phys. Rev. D {\bf
81}, 084035 (2010); X. Calmet, B. Jurco, P. Schupp, J. Wess, and M.
Wohlgenannt, Eur. Phys. J. C {\bf 23}, 363 (2002); B. Meli, K.
Passek-Kumericki, J. Trampetic, P. Schupp, and M. Wohlgenannt, Eur.
Phys. J. C  {\bf 42}, 483 (2005); M. Haghighat and F. Loran, Mod.
Phys. Lett. A {\bf 16}, 1435 (2001); I. Hinchliffe, N. Kersting, and
Y. L. Ma, Int. J. Mod. Phys. A {\bf 19}, 179 (2004); M. Haghighat
and M. M. Ettefaghi, Phys. Rev. D {\bf 70}, 034017 (2004);
 M. M. Ettefaghi and M. Haghighat, Phys. Rev. D {\bf 75}, 125002 (2007);
 C. P. Martin, D. Sanchez-Ruiz, and C. Tamarit, J. High Energy Phys.
 {\bf 02}, 065 (2007);
M. Haghighat, S. M. Zebarjad, and F. Loran, Phys. Rev. D {\bf 66},
016005 (2002); M. Haghighat and F. Loran, Phys. Rev. D {\bf  67},
096003 (2003); M. Chaichian, M. M. Sheikh- Jabbari, and A. Tureanu,
Phys. Rev. Lett. {\bf 86}, 2716 (2001); A. Stern, Phys. Rev. Lett.
{\bf 100}, 061601 (2008);
 P. Schupp, J. Trampetic, J. Wess, and G. Raffelt, Eur. Phys. J. C {\bf 36},
 405 (2004); H. Grosse and Y. Liao, Phys. Rev. D {\bf 64}, 115007 (2001);
 H. Grosse and Y. Liao, Phys. Lett. B {\bf 520}, 63 (2001).

\bibitem{1tev}
M. Haghighat, Phys. Rev. D {\bf 79 }, 025011 (2009); M. M. Ettefaghi
, Phys Rev. D {\bf 79 }, 065022 (2009); R. Horvat and J. Trampetic,
Phys. Rev. D {\bf 79 }, 087701 (2009); A. Joseph,  Phys. Rev. D {\bf
79 }, 096004  (2009); M. Haghighat, M. M. Ettefaghi, and M.
Zeinali, Phys. Rev. D {\bf 73}, 013007 (2006); M. M. Ettefaghi and
M. Haghighat, Phys. Rev. D {\bf 77}, 056009 (2008); Mansour
Haghighat, Nobuchika Okada and Allen Stern, Phys. Rev. D {\bf 82},
016007 (2010); M. Mohammadi Najafabadi, Phys. Rev. D {\bf 74},
025021 (2006); M. M. Ettefaghi, Phys. Rev. D {\bf 86}, 085038
(2012); Weijian Wang, Jia-Hui Huang, and Zheng-Mao Sheng, Phys. Rev.
D {\bf 86}, 025003 (2012);  G. Deshpande, and Sumit K. Garg, Phys.
Lett. B {\bf 708}, 150 (2012); Horvat, A. Ilakovac, P. Schupp, J.
Trampetic, and J. You, JHEP {\bf 1204}, 108 (2012).

\bibitem{ncth}L. Bonora and M. Salizzoni,  Phys.Lett. {\bf B504}, 80-88(2001);
 C. P. Martin and D. Sanchez-Ruiz, Phys. Rev. Lett. {\bf 83}, 476 (1999):
 M. M. Sheikh-Jabbari, JHEP  {\bf 9906}, 015 (1999);
 T. Krajewski and R. Wulkenhaar, Int. J. Mod. Phys.  {\bf A15}, 1011
 (2000); S. Minwalla, M. Van Raamsdonk and N. Seiberg, JHEP {\bf 0002}, 020
(2000); A. Matusis, L. Susskind and N. Toumbas, JHEP {\bf 0012}, 002
 (2000); M. Hayakawa, Phys.Lett. {\bf B478}, 394 (2000);
 A. A. Bichl, J. M. Grimstrup, L. Popp, M. Schweda and R. Wulkenhaar,
Int.J.Mod.Phys. {\bf A17} 2219 (2002); B. Jurco, S. Schraml, P.
Schupp and J. Wess, Eur. Phys. J. {\bf C17}, 521(2000); M. Buric, D.
Latas and V. Radovanovic, JHEP {\bf 0602}, 046 (2006); M. Buric, V.
Radovanovic and J. Trampetic, JHEP {\bf 0703}, 030 (2007); R.
Wulkenhaar,  JHEP {\bf 0203}, 024 (2002): R. Amorim1, and Everton M.
C. Abreu2, Phys. Rev. D {\bf 80}, 105010 (2009); R. Amorim1, Everton
M. C. Abreu, and W. G. Ramirez, Phys. Rev. D {\bf 81}, 105005
(2010); C. P. Martin, Phys. Rev. D {\bf 86}, 065010 (2012).

\bibitem{NC-SME}
S. M. Carroll, J. A. Harvey, V. A. Kostelecky, C. D. Lane, and T.
Okamoto, Phys. Rev. Lett. {\bf 87}, 141601 (2001); S. Aghababaei,
and M. Haghighat, IJPR. {\bf 11}, No.2, 189 (2011).

\bibitem{exp-higgs}
D. L. Anderson, M. Sher, and I. Turan, Phys. Rev. D {\bf 70}, 016001
(2004).






\end{thebibliography}
\end{document}